\documentclass[proceedings]{JHEP} % 10pt is ignored!

\usepackage{epsfig}			% please use epsfig.
\newbox\mybox
\newcommand\fverb{\setbox\mybox=\hbox\bgroup\verb}
\newcommand\fverbdo{\egroup\medskip\noindent\fbox{\unhbox\mybox}\ }
\newcommand\fverbit{\egroup\item[\fbox{\unhbox\mybox}]}

\font\beeg=cmr17 scaled 1600		% Stylish initials
\newcommand\init[1]{\setbox\mybox=\hbox{{\beeg #1}~}%
		   \noindent\global\hangindent=\wd\mybox\global\hangafter-2%
		   \sc\smash{\llap {\lower 13.2pt \box\mybox}}}
\title{Charmonium + Light Hadron Cross Sections}

\author{T.Barnes, E.S.Swanson, C.Y.Wong\\
$^{tb,cyw}$Physics Division, Oak Ridge National Laboratory, Oak Ridge,
TN 37831 USA\\ 
$^{tb}$Department of Physics and Astronomy, 
University of Tennessee, Knoxville, TN
37996 USA\\
$^{tb}$Institut f\"ur Theoretische Kernphysik der Universit\"at
Bonn,  Bonn, D-53115 Germany\\ 
$^{tb}$Institut f\"ur Kernphysik, Forschungszentrum J\"ulich, 
J\"ulich D-52425, Germany\\
$^{ess}$Department of Physics and Astronomy, University of Pittsburg, 
Pittsburg, PA 
15260 USA\\
$^{ess}$Jefferson Laboratory, Newport News, VA 23606 USA\\ 
(presented by T.Barnes)\\
	Email: \email{barnes@bethe.phy.ornl.gov}}

\conference{Heavy Quark Physics 5, Dubna, Russia, 6-8 April 2000}

\abstract
{In this contribution we summarize experimental information and theoretical
results for the dissociation cross sections of charmonium by light hadrons.
Theoretical predictions for these RHIC-related 
processes differ by orders of magnitude over the 
physically relevant energy range.
The results found by the author and collaborators using a 
constituent interchange 
model, which predicts cross sections in the mb region near threshold, 
are discussed in more detail. 
}

\begin{document} 

\section{Introduction}
Many unusual subjects have been studied in the name of QCD. 
One of the more unusual, which has arisen in the field of heavy ion 
collisions, is the size of
cross sections of charmonia on light hadrons. 
This has attracted attention because of the proposal by 
Matsui and Satz \cite{Mat86} that suppression of $J/\psi$  
production could be used as
a signature for the formation of a quark-gluon plasma.

This 
suggestion,
like many signatures proposed for the quark-gluon plasma, 
is perhaps excessively intuitive. The idea is that a QGP will
screen the linear confining interaction between quarks, 
so that a $c\bar c$ pair produced within a QGP will be 
less likely to form a bound $c\bar c$ charmonium resonance, 
as in Fig.1,
but instead will more likely separate to form open-charm mesons.

\FIGURE{\epsfig{file=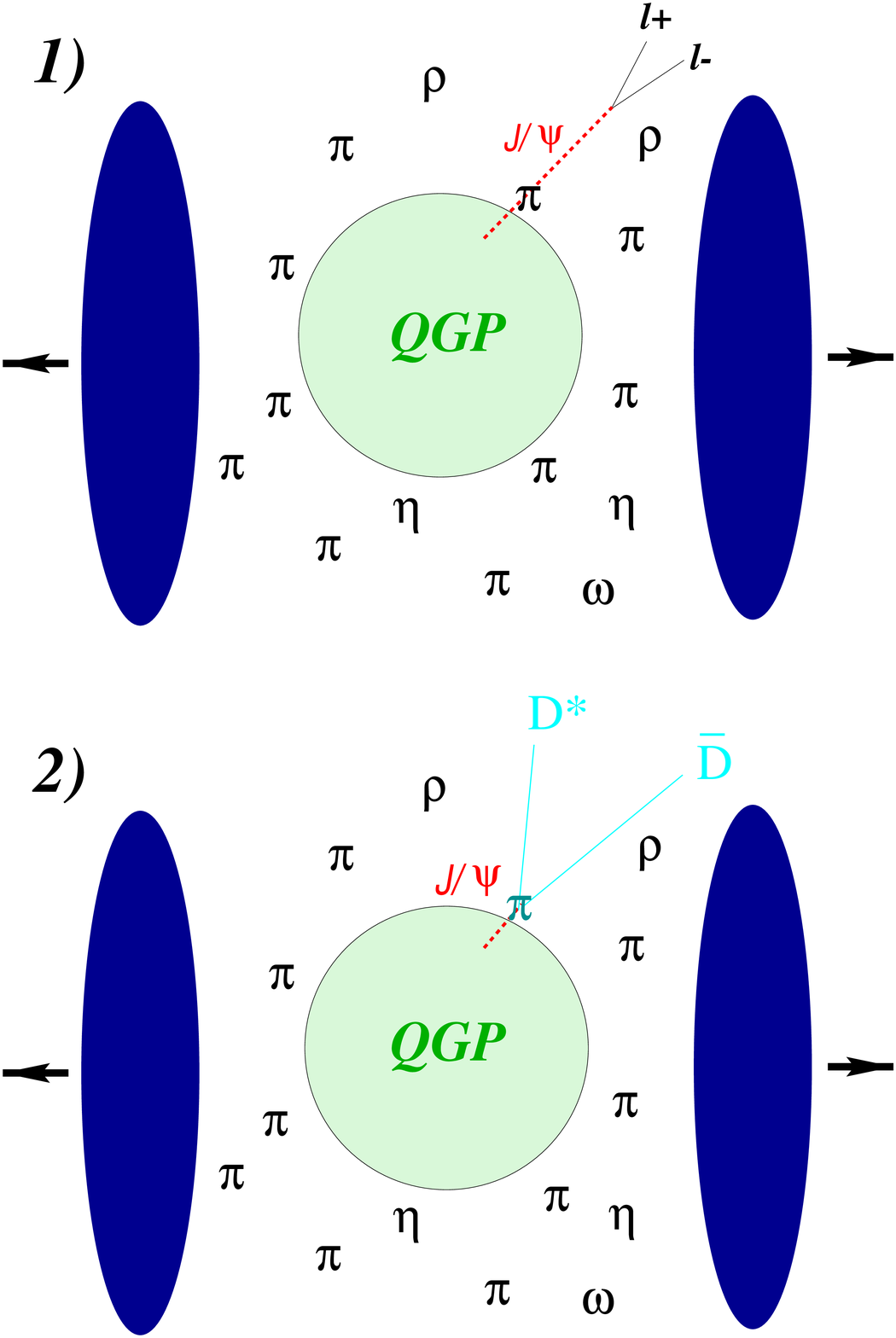,width=5.75cm} 
\caption{The two scenarios we wish to distinguish for the 
external evolution of a
$J/\psi$ produced (purportedly with reduced probability) 
within a quark-gluon plasma: 1) weak, versus
2) strong 
$J/\psi$ absorption by comoving light hadrons.}
\label{dubna1}}

Even if this simple picture of $c\bar c$ production in a QGP 
is qualitatively correct, it can only be confirmed easily if
the competing direct charm production and scattering by the 
initial relativistic nucleons is understood \cite{Hue98a} and 
if there is little subsequent dissociation of the charmonia 
by the many other 
``comoving" light hadrons produced in such a collision. 
To summarize the last point,
if charmonium + light hadron ``comover" 
dissociation cross sections 
are small 
(case 1, top of Fig.1) 
and the 
background of direct 
charm production 
from the initial nucleons is understood,
one may have a useful signature of QGP formation, but 
if the comover dissociation 
cross sections are large
(case 2, bottom of Fig.1) 
one must distinguish 
a QGP-reduced charmonium production amplitude from 
subsequent dissociative scattering, and the interpretation of this 
signal will therefore be ambiguous.

Thus it is of great relevance to the interpretation of 
RHIC physics to establish the 
approximate size of these low energy 
$c\bar c$ + light hadron cross sections.

\section
{Experiment, or what passes for it}
Unfortunately we have no charmonium beams or targets, so the 
experimental cross sections 
must be inferred 
indirectly 
and are poorly known. 
The earliest estimates of lower energy charmonium 
hadronic cross sections
came from $J/\psi$ photoproduction experiments in the
mid 1970s, which were interpreted in terms of a  $J/\psi+N$ 
cross section given additional theoretical assumptions.
Early Fermilab and SLAC photoproduction 
experiments gave 
rough estimates of $\sim 1$ mb for
$\sigma_{J/\psi+N}$, 
assuming vector dominance, for 
photon energies from $E_\gamma \approx 13$ to $200$ GeV \cite{Kna75,Cam75}.
A subsequent SLAC photoproduction 
experiment in 1977 used the $A$ dependence of
$J/\psi$ absorption to infer a rather larger cross section 
of
$\sigma_{J/\psi+N} = 3.5 (0.8)$ mb
at $E_\gamma \approx 17$ GeV
($\sqrt{s} \approx 6$ GeV) 
\cite{And77}. 
The vector dominance hypothesis may have lead to an underestimate of
the cross section in the earlier references \cite{Hue98b}. 

In heavy ion collisions these cross sections may be estimated from
the ratio of lepton pairs produced in the 
$J/\psi$ peak to ``background" Drell-Yan pairs nearby in energy. 
Since the  $J/\psi$ must reach the exterior of the nuclear target 
to decay into a sharp mass
peak, this ratio gives us an estimate of the 
absorption cross section 
through a classical survival probability formula,
\begin{displaymath}
\sigma(J/\psi\to \mu^+\mu^-) / \sigma(Drell-Yan\ \mu^+\mu^-) 
\end{displaymath}
\begin{equation}
= 
\exp(-\rho\; \sigma_{J/\psi + N}^{abs}\; L ) 
\end{equation}
where $\rho$ is the mean nucleon density and $L$ is the estimated 
mean path length in
the experimental nuclear system.
A ``naive" interpretation of the 
$J/\psi$ production data from collisions of various nuclear species
using this formula
gives   
$\sigma_{J/\psi + N}^{abs}\approx 6$ mb at 
$\sqrt{s}\approx 10$ GeV
\cite{Cap97}, with a numerically similar result for the $\psi'$. 
\FIGURE{\epsfig{file=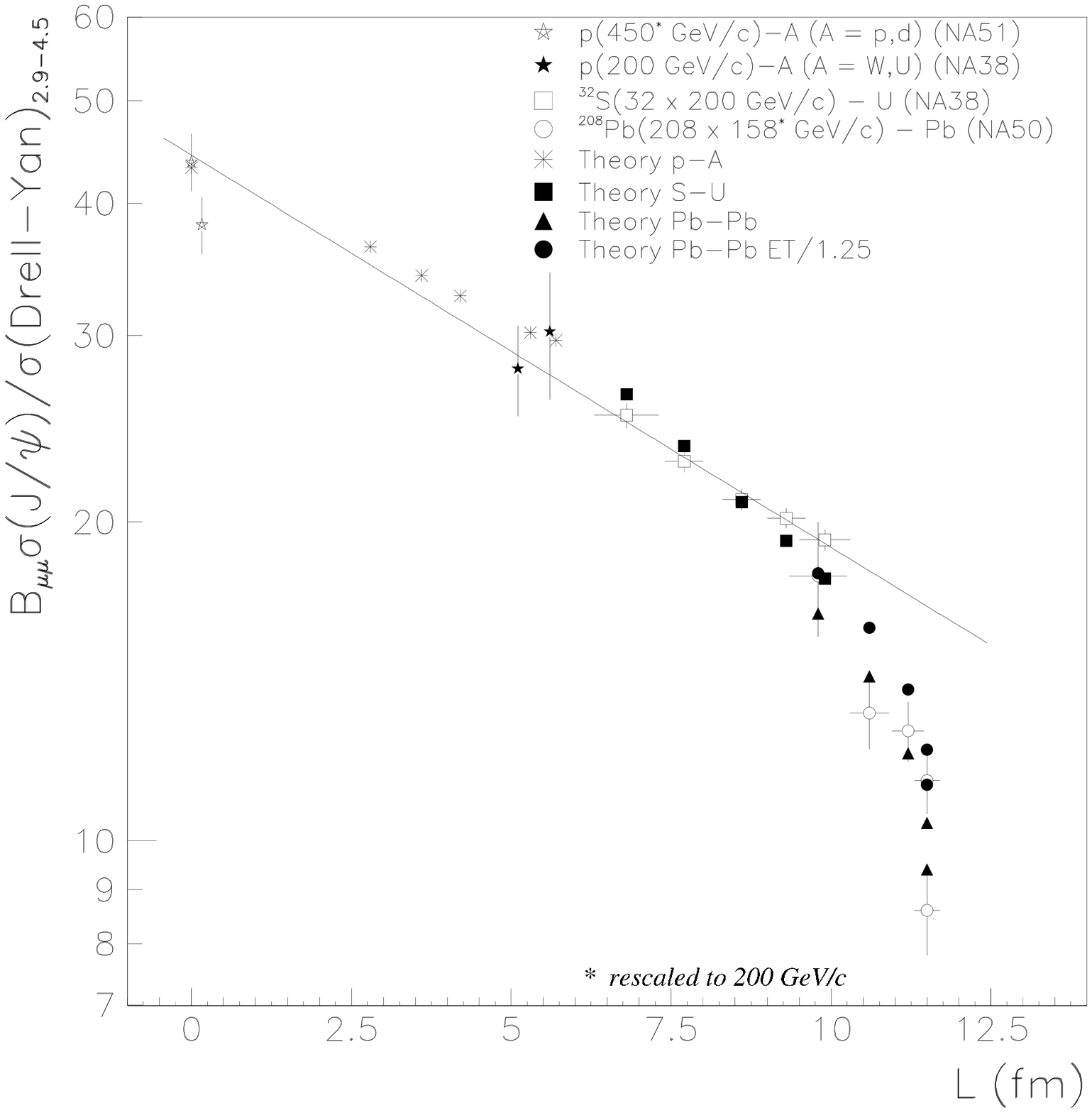,width=7cm} 
\caption{A fit of (2.1) to experimental $J/\psi$ production versus
path length \cite{Cap97}; the line corresponds to 6.2 mb.}
\label{dubna2}}
Of course one may raise many questions about the validity of
this simple estimate,
such as the importance of shadowing in Drell-Yan, the neglect of
$J/\psi$ scattering by other light hadrons formed in the collision
(such as $\pi$ and $\rho$), and the assumption of 
a single, constant $J/\psi+N$ cross section in all circumstances.

Recently, concerns have been expressed that the 
$J/\psi$ and
$\psi'$ 
wavefunctions have not had sufficient time to form within the 
nucleus in these collisions, so experiment may instead be measuring
the cross section for a small initial $c\bar c$ ``premeson" on a nucleon.
One can increase the time spent in the interior of the nuclear system
by selecting small 
and even negative $x_F$ events, as has been done by E866 at Fermilab. 
As discussed by 
He, H\"ufner and Kopeliovich
\cite{Hue96}, this leads one to infer 
$\sigma_{J/\psi + N}^{abs} = 2.8 (0.3)$ mb
and
$\sigma_{\psi' + N}^{abs} = 10.5 (3.6)$ mb respectively, also at 
$\sqrt{s}\approx 10$ GeV.
This is rather more satisfying to people who have an 
intuitive notion that the larger $\psi'$ should have a 
larger reaction cross section. Actually the connection between 
cross section and physical extent is less direct (compare
$KN$ and $\bar KN$),
and in any case the relative proximity of 
inelastic thresholds alone would suggest a 
larger $\psi'$ cross section. These experiments also indicate a 
preference for dissociation over elastic cross sections 
in this energy region by roughly a factor of 30 \cite{Hue96}.

\section{
Theory: Introduction and pQCD}

To quote B.M\"uller, 
``...the state of the theory of interactions 
between $J/\psi$ and light hadrons is embarrassing []. 
Only three serious calculations exist (after more than 10 years of 
intense discussions about this issue!) and their results differ 
by at least two orders of magnitude in the relevant energy range 
[]. There is a lot to do for those who would like to make a 
serious contribution to an important topic." \cite{Mue99}.

The theoretical situation has improved in the subsequent year, 
at least in terms of the quantity of calculations if not in 
the understanding of the scattering mechanism. 
A list of $c\bar c$ + light hadron cross section calculations 
is given in Table I.

The most cited work, {\it albeit} furthest in its predictions from 
a low-energy 
``theoretical mean", is the pQCD calculation of Kharzeev and Satz
\cite{Kha94}. 
This reference is basically a restatement of the 
color-dipole scattering model
developed in the late 1970s by Peskin and Bhanot \cite{Pes79}
for scattering of light hadrons by 
Coulombic bound states of very massive quarks. 
According to 
Peskin, the criterion for validity of this approach is
``...not met even for the $b\bar b$ 
system." 
\cite{Pes00}, so there may be large systematic errors at the
$c\bar c$ mass scale.
This approach certainly makes marginal approximations 
for charmonium, such as the use of Coulombic wavefunctions 
(which are far from accurate for $c\bar c$) and the
introduction of a 
$Q\bar Q$ binding 
energy (which is hard to interpret for charmonium, and 
is 
taken to be $2\, M_D  - M_{J/\psi}$ by Kharzeev and Satz). 
These color-dipole scattering 
formulas also implicitly assume that charmonia are small 
relative to
the natural QCD length scale. 
Since potential 
models actually find rms $c\bar c$ separations 
of about 0.4~fm for the $J/\psi$, 0.6~fm for the $\chi_c$ states and
0.8~fm 
for the $\psi'$ \cite{Buc81}, this is also a dubious assumption. 

\begin{table}
\begin{tabular}{|l||l|c|}
\hline
Method & Init. State & Ref. \cr
\hline 
\hline 
pQCD  & 
$J/\psi + N $ & 
\cite{Kha94}  
\cr
\hline 
meson ex.   & 
$J/\psi + \pi $ & 
\cite{Bla00}  \cr
                 & 
$J/\psi + \pi, \rho $ & 
\cite{Mat98}  \cr
                 & 
$J/\psi + \pi, \rho $ & 
\cite{Lin99}  \cr
                 & 
$J/\psi + N $ & 
\cite{Sib00}  \cr
                 & 
$J/\psi + \pi, K, \rho, N $ & 
\cite{Hag00a,Hag00b}  \cr
\hline 
constit. int.  & 
$J/\psi + \pi$ & 
\cite{Bla00,Mar95} \cr
  & 
$J/\psi + \pi, \rho \; ;\  \psi' + \pi, \rho$ &
\cite{Won99} \cr
  & 
$J/\psi + \pi, N \; ;\  \psi' + \pi, N$ &
\cite{Mar96} \cr
\hline
\end{tabular}
\caption{A summary of $c\bar c$ + light hadron cross section calculations.}
\end{table}

Although this approach has problems with justification  
for $c\bar c$, the predictions are 
nonetheless interesting as estimates of the scale of these cross
sections assuming a color-dipole scattering mechanism, 
and the approach could presumably be extended to lower energies 
by generalizing the
wavefunctions and interaction. 
The formula for the $J/\psi+N$ cross section 
quoted by Kharzeev and Satz \cite{Kha94} is
\begin{equation}
\sigma_{J/\psi+N} = 2.5\hbox{ mb}\ \bigg(1 - {\lambda_0 \over \lambda} 
\bigg)^{6.5} \ ,
\end{equation}
where $\lambda = (s-M_{J/\psi}^2 - M_N^2) / 2\, M_{J/\psi}$,
the constant
$\lambda_0$ is defined to be 
$ `` \simeq  M_N + \epsilon_0 "$ according to the text following 
Eq.(24) of \cite{Kha94} (we assume the equality), 
and the ``binding energy" $\epsilon_0$ is set to $2\, M_D  - M_{J/\psi}$.
The result
is shown in Fig.3, together with the 
single lower-energy SLAC experimental point 
\cite{And77}. 
\FIGURE{\epsfig{file=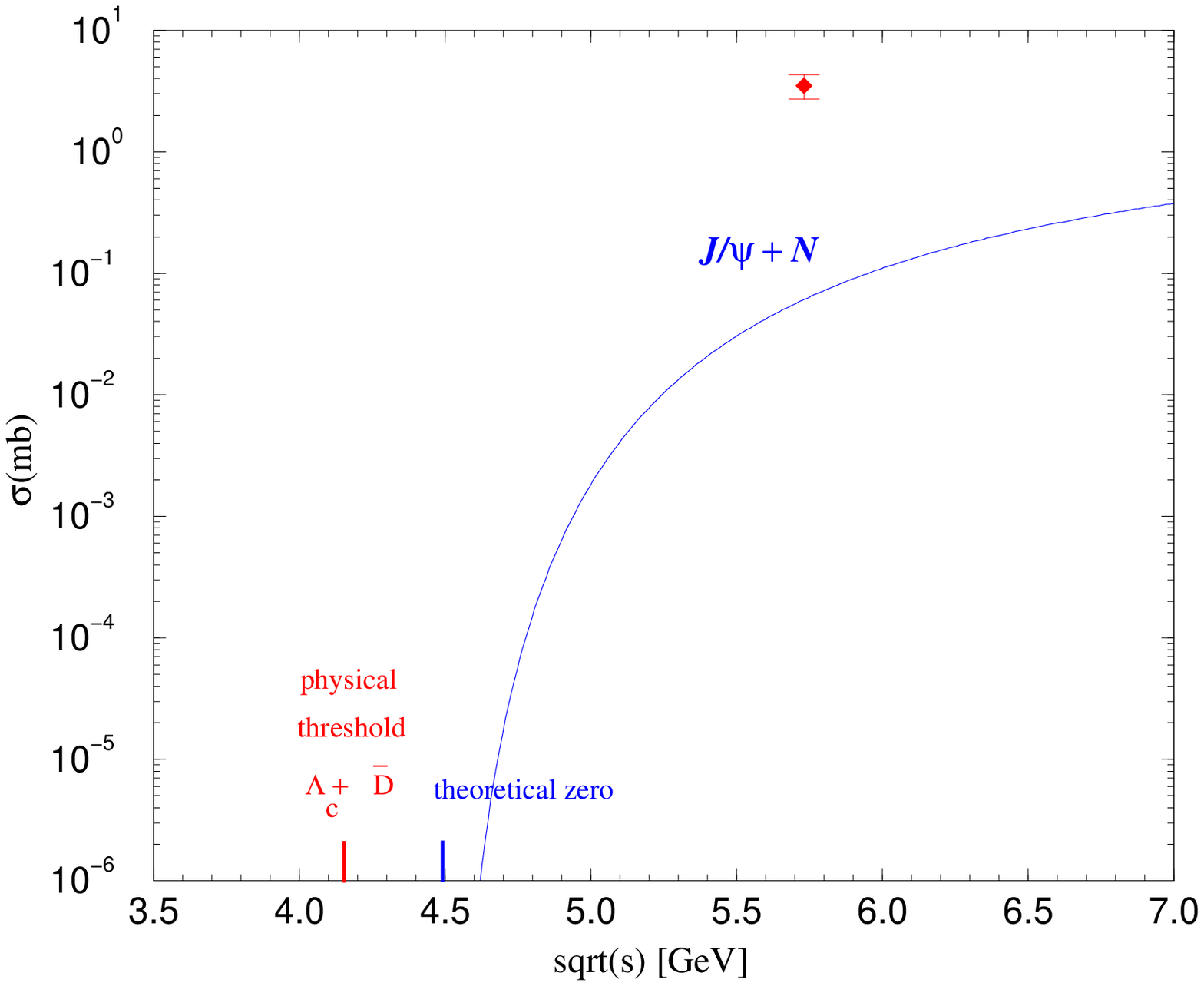,width=7cm} 
\caption{The Kharzeev-Satz $J/\psi+N$ total cross section 
Eq.(3.1) and the 1977 SLAC result \cite{And77}.}
\label{dubna3}}

Evidently the Kharzeev-Satz cross section is smaller than this SLAC 
point (which was an inferred cross section and certainly needs 
confirmation) by about two orders
of magnitude, and falls precipitously as $\sqrt{s}$ is decreased.
This 
approach 
actually has no direct information about physical thresholds, 
so Kharzeev {\it et al.} typically leave their curves ``dangling" just
below 
$\sqrt{s} = 5$ GeV. (See Fig.2 of \cite{Kha94} for example.) 
If we actually plot their formula (3.1) for 
$J/\psi+N$ at low energy (Fig.3), we find the unphysical prediction of a 
zero cross section near 4.5~GeV, whereas the physical 
$\Lambda_c + \bar D$ threshold is at 4.15~GeV. Obviously the method is 
inapplicable at low energies, which is unfortunate 
because this regime is of greatest
interest for RHIC studies of QGP production.
Cross section calculations 
using this approach with more realistic $c\bar c$
wavefunctions are currently being carried out 
by Kopeliovich {\it et al.}
\cite{Kop00}.

\section{
Theory: meson exchange}

Recently, 
several calculations of charmonium + light hadron
cross sections 
have been reported
assuming $t$-channel charmed meson exchange. 
Of course this picture is also problematic, since the 
range of the exchanged charmed meson would be only about $1/M_D \approx 0.1$ fm,
and the assumption of nonoverlapping hadrons at this separation 
is clearly invalid. 
(This is the Isgur-Maltman \cite{Isg84} argument as to 
why vector meson exchange is
unjustifed as the source of the short ranged NN core interaction.) 
Nonetheless it is again interesting to see what scale of 
cross section is predicted by this type of model, 
since it might at least incorporate the correct scales and 
degrees of freedom, and it assumes a different scattering mechanism 
from the pQCD color-dipole model advocated by Kharzeev and Satz.

The first such meson exchange calculation, due to Matinyan and M\"uller
\cite{Mat98}, considered $t$-channel $D$ exchange as the mechanism 
for the reactions
$J/\psi+\pi\to D^*\bar D + h.c.$ and $J/\psi+\rho\to D\bar D$. 
The
cross sections found by this reference 
are shown in Fig.4 below. Note that 500~MeV 
above threshold these cross sections lie in the 0.5 to 1 mb range. 
Similarly, a recent $D$ and $D^*$ meson exchange calculation of 
$J/\psi+N\to \Lambda_c+\bar D$ by Sibirtsev, Tsushima and Thomas
\cite{Sib00} finds a peak cross section of about 2 mb near
$\sqrt{s}=4.6$ GeV. (An earlier calculation of $J/\psi+N\to \Lambda_c+\bar D$ 
by Haglin \cite{Hag00a}, assuming $D$ exchange but no 
hadronic form factors, found
a somewhat larger peak of about 7 mb near $\sqrt{s}=4.3$ GeV.) 
In comparison,
Kharzeev and Satz predict
nanobarn-scale $J/\psi+N$ cross sections 500~MeV above threshold, 
{\it six orders of magnitude 
smaller!} The scatter of theoretical predictions 
for these processes is remarkably wide. 

This 
discrepancy between theoretical cross sections
may appear discouraging. One can instead consider it an opportunity
to learn something important about QCD:
the predictions differ because they 
come from very 
different assumptions regarding the scattering mechanism, and 
as they are
so far apart,
for once we have a clear possibility of 
distinguishing between different 
hadron-hadron scattering models.

\FIGURE{\epsfig{file=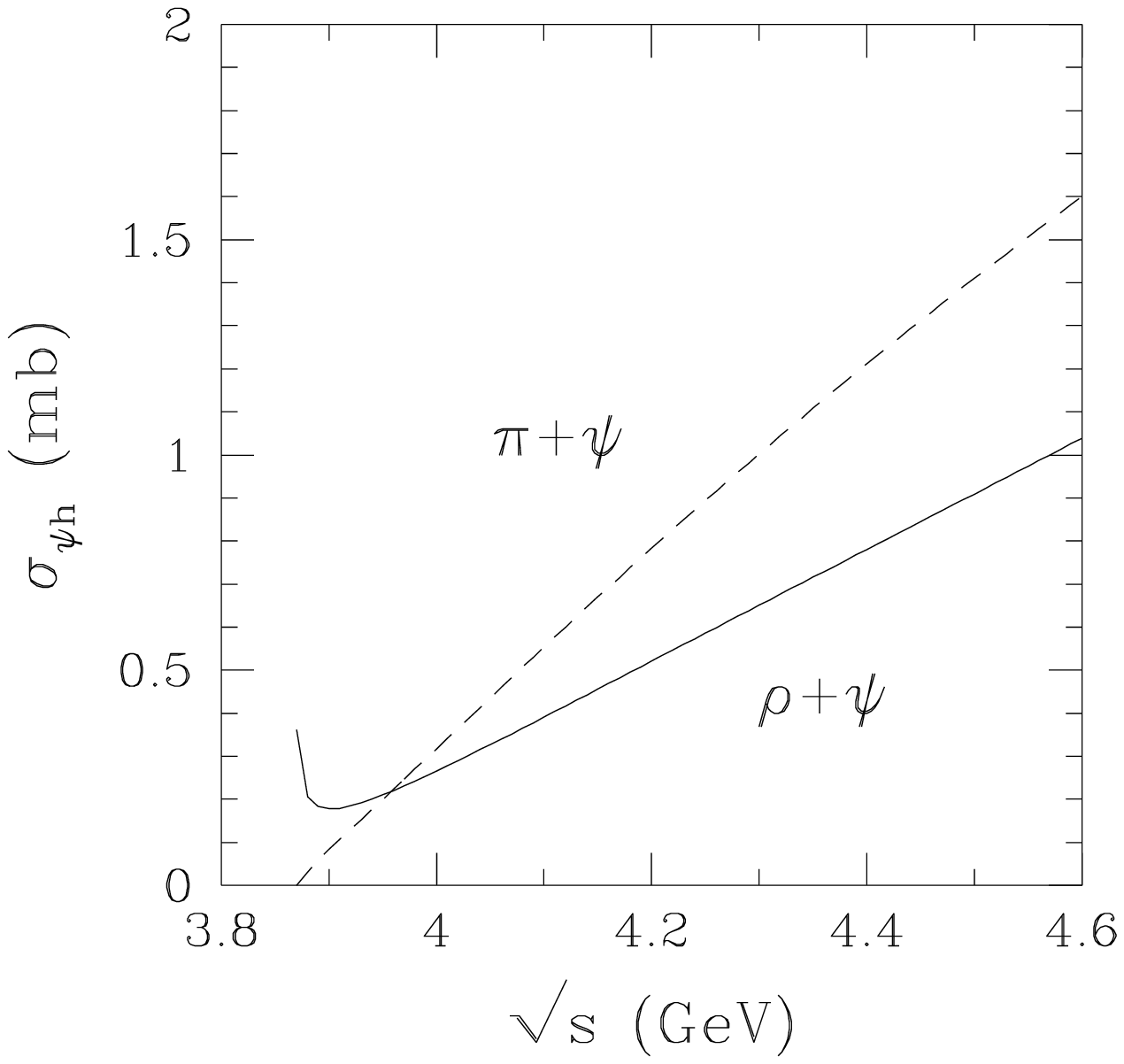,width=7cm} 
\caption{The Matinyan-M\"uller $t$-channel meson exchange results for
$J/\psi+\pi$ 
and
$J/\psi+\rho$ inelastic cross sections 
\cite{Mat98}.}
\label{dubna4}}

\FIGURE{\epsfig{file=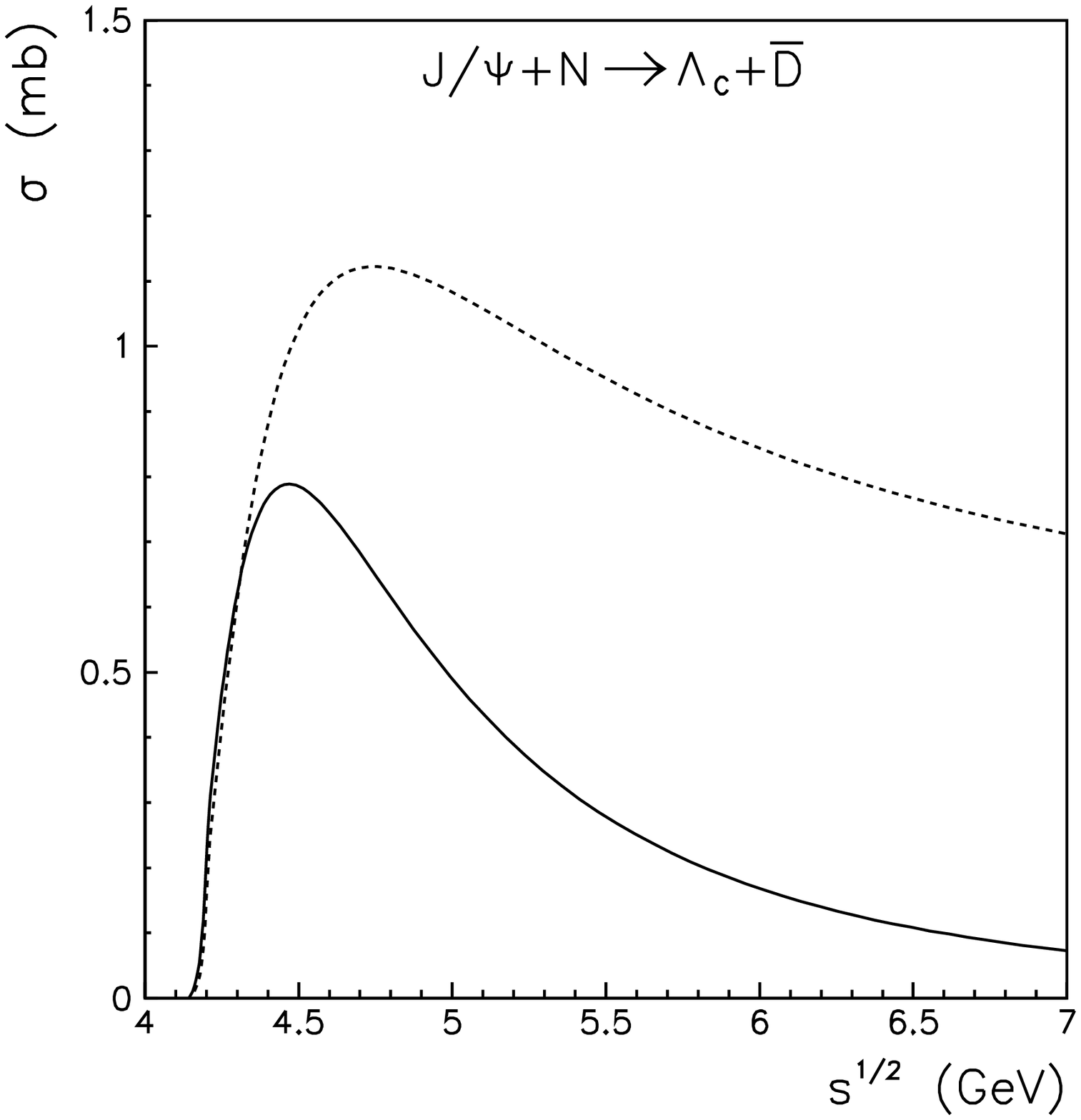,width=7cm} 
\caption{The $t$-channel meson exchange cross sections
for $J/\psi+N \to \Lambda_c+\bar D$ 
found by Sibirtsev, Tsushima and Thomas 
\cite{Sib00}. 
The smaller contribution is from $D$ exchange 
and the larger is from (non-interfering) $D^*$ exchange.}
\label{dubna5}}

Meson exchange 
calculations of $c\bar c$ + $q\bar q$ dissociation cross sections 
have since 
been reported by Lin and Ko \cite{Lin99}
and Haglin and Gale \cite{Hag00a,Hag00b}. 
The models
differ considerably in detail, due to different choices for the
diagrams included, the effective meson Lagrangian used,
and the coupling constants and form factors assumed. 
With pointlike ``hard" vertices, $J/\psi+\pi$ cross sections 
in the 10s of mb not far above threshold are typical. 
After introducing plausible hadronic form factors, 
these are usually reduced to peak values of 
$1$-$10$ mb near  
$\sqrt{s}\approx 4.0$-$4.2$ GeV.
Unfortunately there is 
considerable ``guessing" regarding
hadronic couplings constants, 
which may be unnecessary because these can be calculated 
using well established quark model techniques, 
for example the $^3$P$_0$ meson decay model. Similarly, 
hadronic form factors can be derived from the $^3$P$_0$ model, 
presumably with sufficient accuracy for these purposes.

\section{
Theory: constituent interchange}

We advocate an approach which uses 
nonrelativistic 
quark model wavefunctions and calculates these cross sections 
assuming a constituent interchange scattering 
mechanism, driven by the Born-order matrix element of
the standard quark model Hamiltonian.
This technique, which has no free parameters 
once quark model wavefunctions
and the interquark Hamiltonian are specified,
has been shown to compare reasonably well with
experimental 
low energy hadron-hadron scattering data near threshold
in a wide range of annihilation-free reactions \cite{Bar92,QBD}.
In meson-meson scattering there are four distinct diagrams (see Fig.6),
each of which has an associated overlap integral of the nonrelativistic
quark model external meson wavefunctions convolved with the interquark
Hamiltonian. Constituent interchange is forced at Born-order 
because $H_I\propto \lambda^a \cdot \lambda^a $ changes each
initial color-singlet $q\bar q$ meson into a color octet, which has overlap
with color-singlet final state mesons only after quark line interchange.
The Feynman rules for these diagrams were given by Barnes and Swanson
\cite{Bar92}.

\FIGURE{\epsfig{file=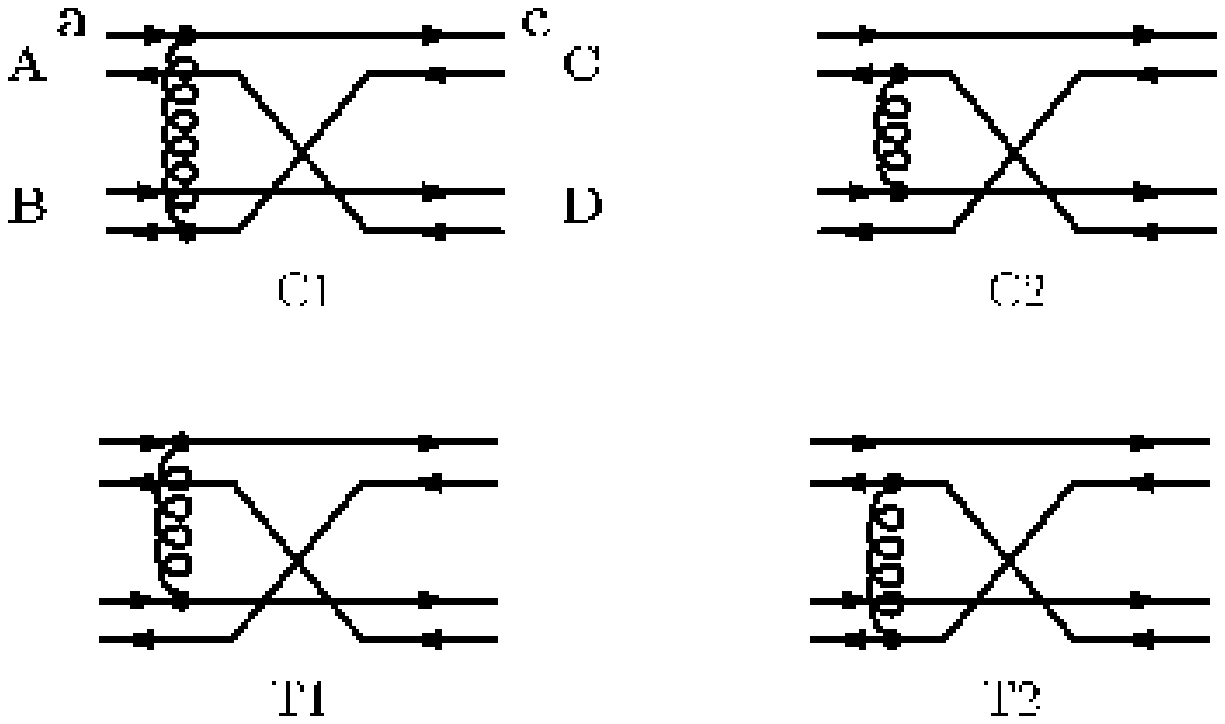,width=7cm} 
\caption{The four constituent interchange scattering diagrams evaluated in the
$J/\psi$ + $q\bar q$ cross section calculation
\cite{Mar95,Won99,Mar96}. The ``exchange" is the full 
quark-quark interaction Hamiltonian $H_I$.}
\label{dubna6}}

The first quark model calculation using this approach was reported by 
Martins, Blaschke and Quack 
\cite{Mar95}, 
who considered the reactions
$J/\psi+\pi \to D^* \bar D +h.c.$ 
and 
$D^* \bar D^* $ (The amplitude for $J/\psi+\pi\to D\bar D$ is zero in 
the nonrelativistic quark model
without spin-orbit forces, and has been found to be quite weak in a relativized
calculation \cite{Bla00}.) 
Martins {\it et al.} found that these exclusive final states have 
numerically rather similar cross sections 
(except for their different thresholds), 
and give a 
maximum total cross section of about 7 mb at 
$\sqrt{s}\approx 4.1$ GeV.
A constituent interchange 
calculation of $J/\psi + N$ and $\psi' + N$ cross sections 
using a simplified quark+diquark model of the nucleon \cite{Mar96} also
found several-mb peak cross sections not far above threshold.

Our collaboration recently 
carried out quite similar constituent interchange quark model
calculations (Wong {\it et al.} \cite{Won99}).
We used numerically determined Coulomb plus linear plus hyperfine 
quark potential model
wavefunctions, and 
evaluated the 
Born-order
meson-meson scattering amplitude, which is 
the matrix element of the interquark Hamiltonian 
between scattering states with quark interchange (Fig.6). 
We included smeared
OGE spin-spin,
OGE color Cou- lomb and linear confining interactions, with parameters 
chosen to give a good fit to the $q\bar q$ and $c\bar c$ meson spectra. 
We  
find a somewhat
smaller cross section (peaking at about 1~mb at
$\sqrt{s} \approx 4.0$ GeV) for the sum 
of these inelastic $J/\psi+\pi$ reactions. 
The difference between our work and that of
Blaschke {\it et al.} 
lies mainly in the treatment 
of the confining interaction; for simplicity, 
Blaschke {\it et al.} treated confinement
as a 
color-independent Gaussian potential that acts only between 
quark and antiquark (hence they include only diagrams C1 and C2), 
whereas we used the 
standard 
$\lambda^a \cdot \lambda^a$ 
linear 
confining potential 
between all pairs of constituents. 

\vskip -2cm
\FIGURE{\epsfig{file=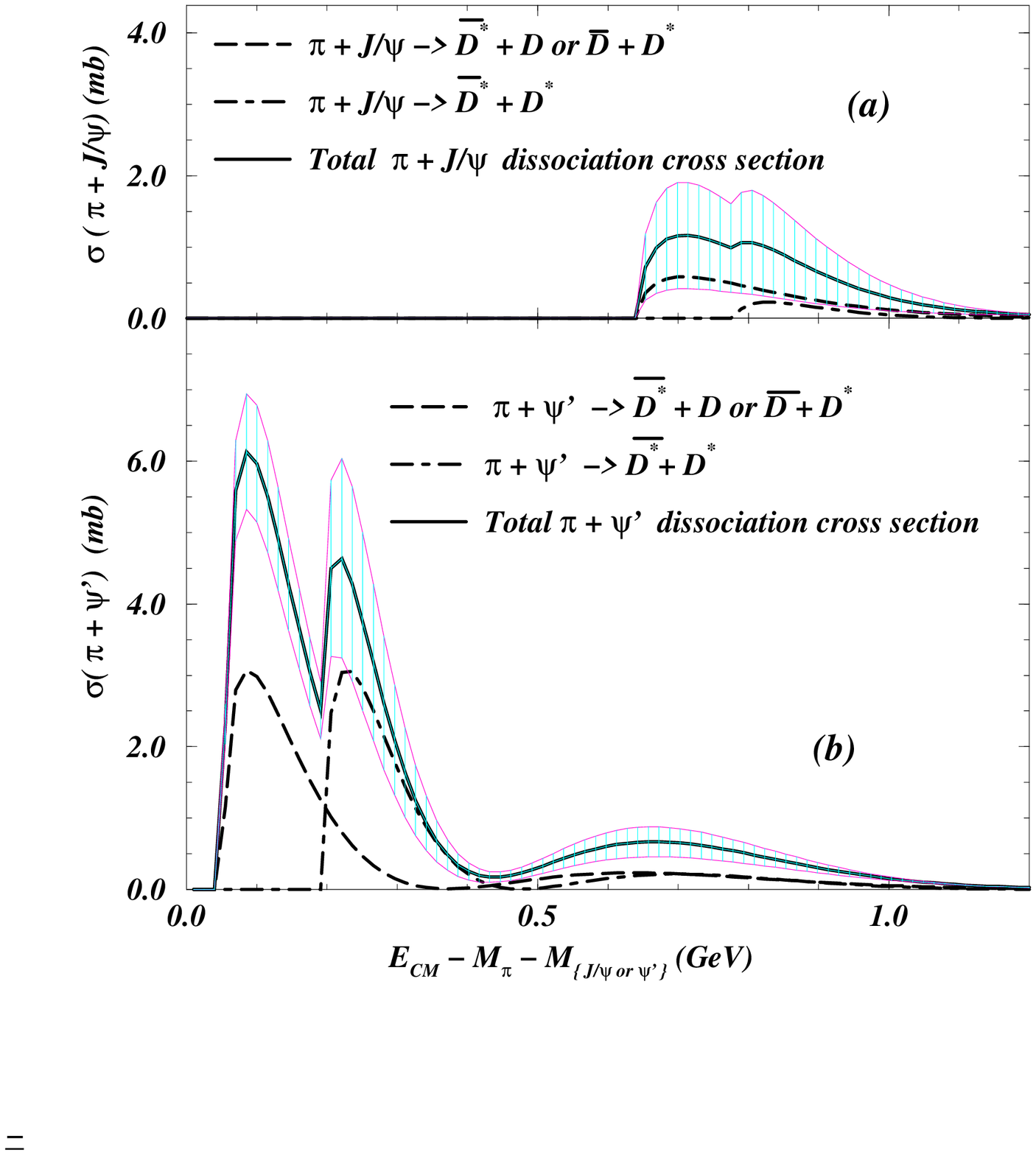,width=7cm} 
\caption{Constituent interchange model predictions for 
$J/\psi+\pi$ and
$\psi'+\pi$ 
cross sections
\cite{Won99}. A band of estimated systematic uncertainty 
is also shown.}
\label{dubna7}}

We find 
destructive interference between the C and T diagrams, 
leading to a much reduced 
total cross section. In \cite{Won99} we also treated $\psi'+\pi$ scattering, 
which involved a simple change to a 2S 
$c\bar c$ wavefunction and a change of phase space,
and found a rather 
large, {\it ca.} 5~mb maximum cross section (see Fig.7). 
Our $c\bar c$ + $q\bar q$ cross sections have 
their strongest support just a few hundred MeV 
in $\sqrt{s}$ above threshold, 
since the overlap integrals are damped by the tails of the 
wavefunctions at higher energies. 

\FIGURE{\epsfig{file=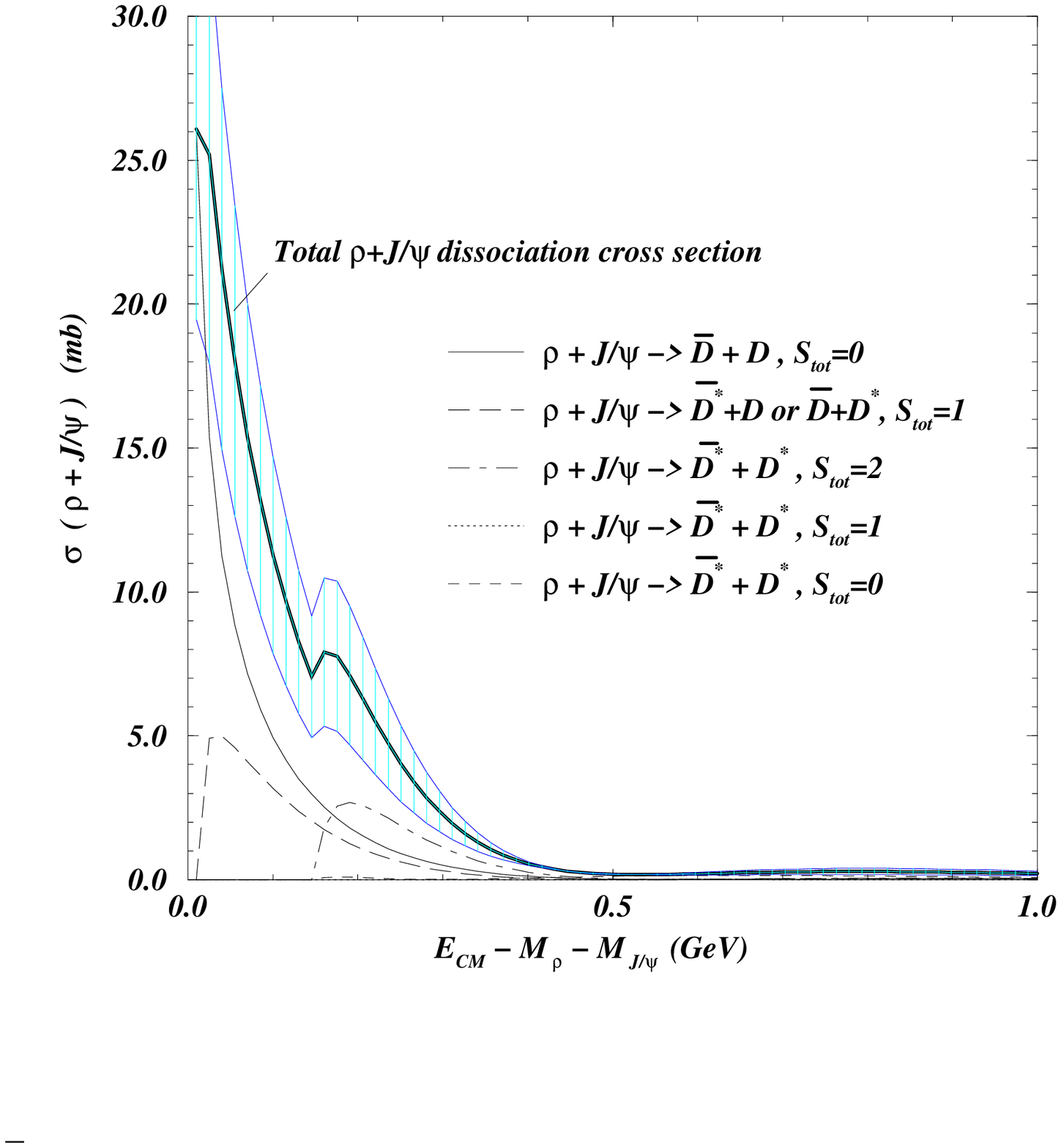,width=7cm} 
\caption{Constituent interchange model predictions for 
$J/\psi+\rho$ 
exothermic reactions
\cite{Won99}, as in Fig.7.}
\label{dubna8}}

We also 
consider $J/\psi+\rho$ (Fig.8) 
and $\psi'+\rho$, which are predicted to be much 
larger near threshold for the simple reason that they are exothermic;
there is a $1/v_{AB}$ divergence in these cross sections 
as we approach threshold.

Although our scattering 
amplitudes and cross sections 
are evaluated numerically, it is interesting that a simple
two-parameter 
function gives a useful approximation to our (single channel) 
cross sections. 
This function is
\begin{equation}
\sigma(s) = \sigma_{max} \; 
(\epsilon / \epsilon_{max} )^p \; 
e^{p(1-\epsilon / \epsilon_{max} )}  \ ,
\end{equation}
where $\epsilon =\sqrt{s} - M_C - M_D$, and $\sigma_{max}$ (mb) is the maximum
value of the cross section, at 
$\epsilon_{max}$ (MeV). 
The threshold exponent $p$ is fixed by the angular quantum numbers of
the hadrons, and is
$\pm 1/2 + L_{min}^{CD}$ (for endothermic/exothermic),
where $L_{min}^{CD}$ is the lowest angular momentum
allowed for the final meson pair $CD$. As an example, in Fig.9 we show
our numerical results for the reaction $\eta_c+\pi^+\to D^+\bar D^o$ 
and a fit using the function (5.1) with  $p=1/2$, as appropriate for 
an S-wave-allowed final state. The masses and parameters assumed 
for this example were
$M_{\pi^+} = 0.140$ GeV,
$M_{\eta_c} = 2.98$ GeV,
$M_{D^+} = 1.869$ GeV,
$M_{D^o} = 1.865$ GeV,
$\alpha_s = 0.6$,
$b$ (string tension) $ = 0.18$ GeV$^2$,
$m_{u,d} = 0.33$ GeV,
$m_c = 1.6$ GeV, and the OGE contact hyperfine smearing distance was
$0.25$ fm. These are all reasonably well established 
nonrelativistic quark model parameters. 
The Schr\"odinger equation was solved with these parameters
to generate numerical wavefunctions, which were then used 
in a nine-dimensional Monte Carlo integration
to evaluate the 
scattering amplitudes 
in the CM frame.
The Monte Carlo 
amplitude evaluation in Fig.9 used
$4M$ points for each diagram 
at each energy and each final $D^+(\hat\Omega)$ direction; 
amplitudes along three final
directions were evaluated, 
which were then projected into S-, P- and D-moments to separate the
different partial waves. Cancellation of diagram sums in certain channels
as well as evaluation of known exact results with SHO wavefunctions
provided nontrivial checks of the numerical work.

\FIGURE{\epsfig{file=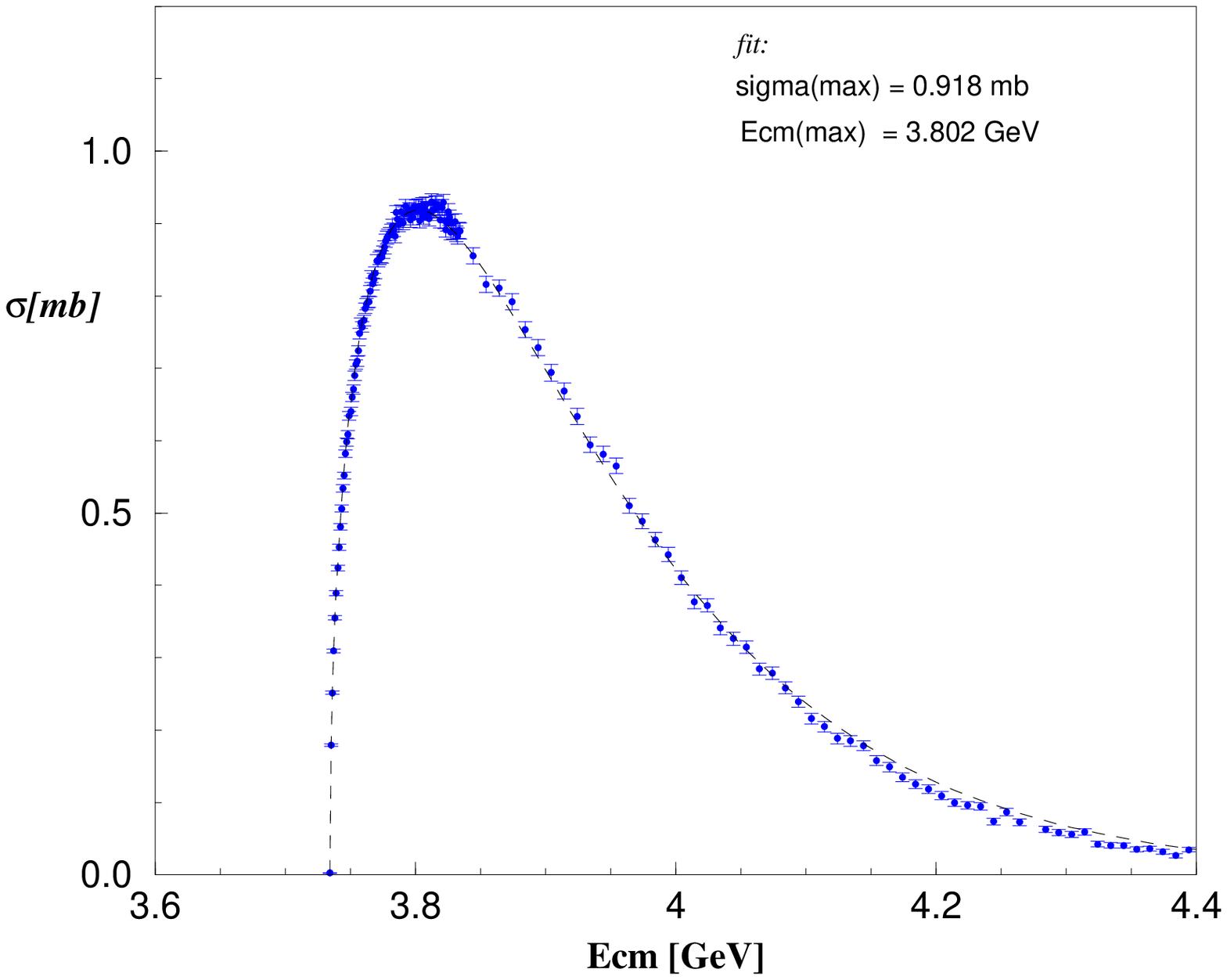,width=7cm} 
\caption{Monte Carlo evaluation of the cross section for
$\eta_c + \pi^+ \to D^+ \bar D^o$ in the constituent interchange model.
A fit to the function (5.1) is also shown.}
\label{dubna9}}

\section{Conculsions}
We have reviewed the recent theoretical predictions and experimental
status of the cross sections of $c\bar c$ on light hadrons, which is of
great interest for the interpretation of heavy ion collisions.
There are three scattering mechanisms currently being 
investigated, which are color-di- pole pQCD, $t$-channel meson exchange, and 
constituent interchange. The pQCD approach gives very small cross sections 
at low energies,
whereas the meson exchange and constituent interchange models
both predict peak cross sections 
near thres- hold of $\approx 1$-$10$~mb. If it is possible to
establish these cross sections experimentally, we may achieve
a much better understanding of the mechanisms of low energy 
hadron-hadron scattering.

\section{Acknowledgements}
We would like to acknowledge the kind invitation and support of the
organisers of ``Heavy Quark Physics 5", which made it possible 
to attend the meeting and present these results.
This research was supported in part by the DOE Division of Nuclear Physics, 
at ORNL,
managed by UT-Bat- telle, LLC, 
under Contract No. DE-AC05-00OR 22725, 
Research Corp, 
and by the Deutsche For- schungsgemeinschaft DFG under contract Bo
56/ 153-1.  We would also like to thank D.B.Blaschke, B.Kopeliovich, M.Peskin,
	A.Sibirtsev and S.Sor- ensen for useful discussions and information.

\end{document}